\documentclass[11pt, a4paper]{article}
\usepackage[backend=biber, style=numeric-comp, sorting=none, natbib=true, url=false]{biblatex}
\usepackage{authblk}
\usepackage{amsmath,amssymb,graphicx,epsfig,latexsym,tensor,verbatim,siunitx}
\usepackage{physics,tensor,braket,xcolor}
\numberwithin{equation}{section}

\title{Tensor extension of the Abelian-Higgs model for a superconductor}

\author{
	Spyros Konitopoulos$^{1,2}$\thanks{\texttt{spykoni@gmail.com}} and Elias Koorambas$^{1}$\thanks{\texttt{elias.koor@gmail.com}}\\\smallskip
	\small $^{1}$Institute of Nuclear and Particle Physics, NCSR Demokritos, Athens, Greece \\
	\small $^{2}$University of Derby, Mediterranean College, Athens, Greece
}
\date{\today}

\begin{document}

\setlength{\baselineskip}{7mm}
\begin{titlepage}
\maketitle
\abstract{We extend the Abelian-Higgs model of superconductivity to incorporate higher-spin particles. Microscopically, these higher-spin states can be modeled as multi-electron clusters, such as spin-1 Copper pairs or quartets, existing alongside the standard Cooper pairs predicted by BCS theory. To account for these composites, we introduce vector and higher-rank tensor non-gauge fields into the Lagrangian, which serve as sources for higher-rank tensor gauge fields. In this work, we extend the particle spectrum by one rank (including the necessary auxiliary fields) and examine the resulting modifications to the fundamental phenomenological parameters of superconductivity, specifically the penetration depth and the correlation length.}
\end{titlepage}
\newpage
\pagestyle{plain}

\section{Introduction}
The Abelian Higgs model serves as a relativistic analogue to the Ginzburg-Landau (GL), phenomenological model of superconductivity. It consists of a complex scalar field, representing the superconducting condensate, coupled to the $U(1)$ electromagnetic, gauge field. In this context, the Higgs mechanism corresponds to the Meissner effect, which provides mass to the photon and expels the magnetic field from the bulk of the superconductor. \cite{Ginzburg:1950,suzuki2012,Fraser2016,Gros,Poniatowski:2019,Tinkham1996,Weinberg1986,Weinberg2005,Ouboter2017}.

The BCS Hamiltonian provides a microscopic, dynamical counterpart to the GL model. It demonstrates that the Fermi sea of electrons becomes unstable toward the formation of singlet Cooper pairs, an effect that plays an essential role in the transition to the superconducting state \cite{BCS,Tinkham1996}. Nevertheless, strong ferromagnetic interactions favor the parallel spin alignment of electrons, which opens the possibility for triplet superconductivity. Strontium Ruthenate ($\text{Sr}_2\text{RuO}_4$) and Uranium Platinum-3 ($\text{UPt}_3$) represent prototypical examples of such unconventional superconductors. In these systems, the instability of the Fermi sea toward Cooper pairing occurs in non-trivial channels \cite{Timm2020}. Furthermore, in certain unconventional superconductors, such as those found in three dimensional systems or twisted bilayer graphene, the pairing is described by a complex tensor order parameter (e.g., Spin-2 Cooper pairs) \cite{Gibney2018}. To investigate these states, researchers have developed methods using superconducting qubits and circuits to experimentally measure the quantum metric tensor, allowing for the direct study of the geometry and topology of quantum states in these systems \cite{Weisbrich2021,Xie2022}.

A comprehensive theory of higher-spin fields has been developed by G. Savvidy \cite{Savvidy2005,Savvidy2006,Savvidy2009,Savvidy2010}. This theoretical framework is based on a natural extension of the gauge invariance principle, which facilitates the inclusion of higher-spin fields across the bosonic, fermionic, and Higgs sectors\footnote{In a previous study, one of the authors explored whether dark matter could be explained by a new, neutral, non-symmetric tensor gauge boson (the $Z_{\mu\nu}$-boson, with a mass of \SI{2.85}{\TeV} \cite{Koorambas2023}) predicted by the tensor gauge extension of the Electroweak (EW) theory \cite{Savvidy2005,Savvidy2006,Savvidy2009,Savvidy2010}.}. Building on this foundation, we investigate whether this model offers a plausible phenomenological extension of the standard Abelian Higgs model to account for unconventional superconductivity.

The current study is organized as follows: Section 2 provides a summary of the Abelian-Higgs model, followed by an exploration of its higher-spin extension in Section 3. In Section 4, we utilize the proposed model to derive tensor supercurrents that are proportional to the massive tensor gauge fields. Furthermore, by leveraging the underlying symmetries of the theory, we identify both symmetric and antisymmetric second-rank tensor magnetic fields, ultimately establishing the corresponding London equations.

\section{Preliminaries}
The Lorentz invariant analogue of the phenomenological Landau-Ginzburg theory for a superconductor \autocite{Ginzburg:1950,Tinkham1996,suzuki2012,Ouboter2017,Gros,Charbonneau2005}, is the Abelian Higgs model \autocite{Fraser2016,Poniatowski:2019}.
In this model, a self interacting scalar is coupled to the electromagnetic field. The full Lagrangian is,
\begin{equation}\label{Standard abelian Higgs}
\mathcal{L}=-\frac{1}{4}F_{\mu\nu}F^{\mu\nu}+(D_\mu\phi)^*D^\mu\phi-\frac{1}{2}\lambda^2\left(\phi^*\phi-\frac{1}{2}\eta^2\right)^2
\end{equation}
where,
\begin{equation}
D_\mu\phi=(\partial_\mu-igA_\mu)\phi
\end{equation}
The Lagrangian \eqref{Standard abelian Higgs} is invariant under $U(1)$ gauge transformations, which in the infinitesimal limit are written,
\begin{eqnarray}
&&\delta\phi=ig\xi\phi\\
&&\delta A_\mu=\partial_\mu\xi
\end{eqnarray}
We get the Noether current if we perform a global $U(1)$ variation on \eqref{Standard abelian Higgs},
\begin{equation}\label{Noether e/m}
j^\mu=\frac{1}{g\xi}\left[\frac{\partial\mathcal{L}}{\partial(\partial_\mu\phi)}\delta\phi+
\frac{\partial\mathcal{L}}{\partial(\partial_\mu\phi^*)}\delta\phi^*\right]=
i\left[\phi(D^\mu\phi)^*-\phi^*(D^\mu\phi)\right]
\end{equation}
The Euler-Lagrange equations for the $A_\mu$ field are,
\begin{equation}
\frac{\partial\mathcal{L}}{\partial A_\mu}=\partial_\nu\left(\frac{\partial\mathcal{L}}{\partial(\partial_\nu A_\mu)}\right)
\end{equation}
We get, 
\begin{equation}
\frac{\partial\mathcal{L}}{\partial A_\mu}=-gj^\mu~~\text{and}~~
\frac{\partial\mathcal{L}}{\partial(\partial_\nu A_\mu)}=F^{\mu\nu}
\end{equation}
Therefore, we extract the manifestly covariant form of the dynamical Maxwell equations\footnote{Here and henceforth, we use the metric $(+,-,-,-)$.},
\begin{equation}\label{Maxwell}
\partial_\mu F^{\mu\nu}=gj^\nu
\end{equation}
The energy of the system acquires a minimum at $|\phi_0|=\frac{\eta}{\sqrt{2}}$, giving an $S^1$ vacuum manifold parametrized by the phase $\theta(x)$. The abelian symmetry is spontaneously broken when a specific vacuum is chosen. Expanding the fields from this vacuum, we set, 
\begin{equation}
\phi(x)=\frac{1}{\sqrt{2}}(\eta+\sigma(x))e^{\theta(x)}
\end{equation}
After the SSB and fixing in the unitary gauge\footnote{In the unitary gauge, we choose $A_\mu\rightarrow A_\mu-\frac{1}{g}\partial_\mu\theta$ and the Goldstone boson disappears completely.} the Lagrangian becomes,
\begin{equation}
\mathcal{L}=-\frac{1}{4}F_{\mu\nu}F^{\mu\nu}+\frac{1}{2}m_\gamma^2A_\mu A^\mu+\frac{1}{2}\partial_\mu\sigma\partial^\mu\sigma-\frac{1}{2}m_\sigma^2\sigma^2+(\text{interaction terms})
\end{equation}
The photon has acquired a mass $m_\gamma=g\eta$ and, therefore, a longitudinal degree of freedom which  has been transferred from the phase of the complex scalar field $\phi$. The surviving degree of freedom has become a real Klein-Gordon field with mass $m_\sigma=\lambda\eta$.

As mentioned, in the ground state the scalar field acquires the vacuum expectation value $\phi_0(x)=\frac{1}{\sqrt{2}}\eta$. Inserting this into \eqref{Noether e/m} and remaining in the unitary gauge, we get the ground state current,
\begin{equation}\label{London e/m}
j^\mu=-g\eta^2A_\mu
\end{equation}
The above equation encompasses the London equations and therefore the basic exotic properties of the superconductor (zero resistivity and exclusion of magnetic fields).  

The masses of the photon and the $\sigma$ field, after the SSB, are of fundamental phenomenological significance for superconductors. The massive photon satisfies a Proca equation\footnote{This is easily seen if \eqref{London e/m} is inserted in \eqref{Maxwell}.} and the inverse of its mass coincides with the London's penetration's depth, 
\begin{equation}
    \lambda_L=\frac{1}{g\eta}
\end{equation}
which measures the microscopic scale an external magnetic field enters the superconductor.  
The inverse of $m_\sigma$ coincides with the correlation length, the characteristic distance over which the Cooper-pair wavefunction can vary significantly,
\begin{equation}\label{scalar correlation length}
    \xi_c=\frac{1}{\lambda\eta}
\end{equation}
The energy gap between a normal conductor and a superconductor vacua is \cite{Weinberg2005,Weinberg1986},
\begin{equation}
\Delta=\frac{1}{8g^2\lambda_L^2\xi_c^2}.
\end{equation}

\section{Tensor extension of the Abelian-Higgs model}
Tensor gauge field theory is a natural extension of Yang-Mills theory to account for particles of arbitrary high spin. Gauge bosons of higher integer spin are described by higher rank tensor gauge fields,
\begin{equation}
    A_{\mu\lambda_1\dots\lambda_s}
\end{equation}
on the condition that they are invariant under permutations of the $\lambda_i$ indices \autocite{Savvidy2006}.

Our intention would be to employ abelian tensor gauge theory up to tensors of the 2nd rank, as a natural extension of the Higgs-Anderson model \autocite{Anderson:1963}. Nevertheless, gauge invariance is ensured provided an auxiliary, non-dynamical, 3rd rank tensor gauge field is also included. A scalar $\xi$, a vector $\xi_\mu$ and a 2nd rank $\xi_{\mu\nu}$ gauge functions are introduced, so that that gauge boson fields transform under the $U(1)$ group as follows,
\begin{eqnarray}
&&\delta A_\mu=\partial_\mu\xi\label{gauge transformations_gauge boson_1}\\
&&\delta A_{\mu\nu}=\partial_\mu\xi_\nu\label{gauge transformations_gauge boson_2}\\
&&\delta A_{\mu\nu\lambda}=\partial_\mu\xi_{\nu\lambda}
\label{gauge transformations_gauge boson_3}
\end{eqnarray}
The field strength tensors are accordingly defined as,
\begin{eqnarray}
&&F_{\mu\nu}=\partial_\mu A_\nu-\partial_\nu A_\mu\\
&&F_{\mu\nu,\lambda}=\partial_\mu A_{\nu\lambda}-\partial_\nu A_{\mu\lambda}\\
&&F_{\mu\nu,\lambda\rho}=\partial_\mu A_{\nu\lambda\rho}-\partial_\nu A_{\mu\lambda\rho}
\end{eqnarray}
and are invariant under the gauge transformations \eqref{gauge transformations_gauge boson_1}, \eqref{gauge transformations_gauge boson_2} and \eqref{gauge transformations_gauge boson_3} respectively.

The Lagrangian for the abelian, gauge boson fields will be a linear combination of the common Lagrangian of electromagnetism and of the one that desciribes the kinetic behavior of the 2nd rank tensor gauge field,
\begin{equation}
\mathcal{L}_B=\mathcal{L}_1+g_2\mathcal{L}_2
\end{equation}
where\footnote{The full $\mathcal{L}_2$ Lagrangian exhibits an enhanced gauge symmetry which fixes the number of propagating modes of the 2nd rank gauge field $A_{\mu\nu}$.},
\begin{eqnarray}
\mathcal{L}_1&=&-\frac{1}{4}F_{\mu\nu}F^{\mu\nu}\\
\label{2nd rank bosonic Langrangian}
\mathcal{L}_2&=&-\frac{1}{4}F_{\mu\nu,\lambda}F^{\mu\nu,\lambda}+
\frac{1}{4}F_{\mu\nu,\lambda}F^{\mu\lambda,\nu}+
\frac{1}{4}\tensor{F}{_{\mu\nu,}^\nu}\tensor{F}{^{\mu\lambda,}_\lambda}-
\frac{1}{4}F_{\mu\nu}\tensor{F}{^{\mu\nu,\lambda}}{_\lambda}+
\frac{1}{2}F_{\mu\nu}\tensor{F}{^{\mu\lambda,\nu}}{_\lambda}\nonumber\\
\end{eqnarray}
The arbitrary parameter $g_2$ that has been introduced will end up to be of a high phenomenological importance in the discussion that follows.

To apply the Higgs mechanism, aside from a complex scalar field $\phi$, a complex vector, and a 2nd rank tensor non-gauge fields $\phi_\mu$, $\phi_{\mu\nu}$ are introduced. Altogether, the $\Phi$ fields are subjected to the following gauge transformations \autocite{Savvidy2006},
\begin{eqnarray}
    \delta\phi&=&ig\xi\phi\label{scalar phi gauge transformation}\\
\delta\phi_\mu&=&ig\left(\xi\phi_\mu+\xi_\mu\phi\right)\label{vector phi gauge transformation}\\ \delta\phi_{\mu\nu}&=&ig\left(\xi\phi_{\mu\nu}+\xi_\mu\phi_\nu+\xi_\nu\phi_\mu+\xi_{\mu\nu}\phi\right)\label{tensor phi gauge transformation}
\end{eqnarray}
The symmetry exhibits a local behavior if covariant derivatives $D_\mu$ are introduced, operating on the $\Phi$ fields as follows:
\begin{eqnarray}
    &&D_\mu\phi=\left(\partial_\mu-igA_\mu\right)\phi\label{covariant scalar phi}\\
    &&D_\mu\phi_\nu=\left(\partial_\mu-igA_\mu\right)\phi_\nu\label{covariant vector phi}\\
    &&D_\mu\phi_{\nu\lambda}=\left(\partial_\mu-igA_\mu\right)\phi_{\nu\lambda}\label{covariant tensor phi}
\end{eqnarray}
The gauge transformations of \eqref{covariant scalar phi}-\eqref{covariant tensor phi} are induced directly by
\eqref{scalar phi gauge transformation}-\eqref{tensor phi gauge transformation}. Indeed,
\begin{eqnarray}
    \delta\left(D_\mu\phi\right)&=&ig\xi D_\mu\phi\label{gauge transformation scalar covariant}\\
    \delta\left(D_\mu\phi_\nu\right)&=&ig\left(\xi D_\mu\phi_\nu+\xi_\nu D_\mu\phi+\phi\partial_\mu\xi_\nu\right)\label{gauge transformation vector covariant}\\
    \delta\left(D_\mu\phi_{\nu\lambda}\right)&=&ig\left(\xi D_\mu\phi_{\nu\lambda}+2D_\mu\phi_{(\lambda}~\xi_{\nu)}+\xi_{\nu\lambda}D_\mu\phi+\phi\partial_\mu\xi_{\nu\lambda}+2\partial_\mu\xi_{(\nu}~\phi_{\lambda)}\right)\label{gauge transformation tensor covariant}
\end{eqnarray}
Covariant differentiation of the gauge functions $\xi$, $\xi_\mu$, $\xi_{\mu\nu\lambda}$ coincides with partial differentiation, since we are dealing with an abelian symmetry group\footnote{If the gauge group was non-Abelian, then the covariant derivative in any Lie algebra valued object would add the commutator of the gauge field with that object, e.g. $D_\mu\xi=\partial_\mu\xi-ig[A_\mu,\xi]$.}.

The total Lagrangian for the $\Phi$ fields, including their interactions with the gauge bosons and the potential functional that will determine their vacuum expectation values is, 
\begin{eqnarray}
\mathcal{L}_\Phi&=&(D_\mu\phi)^*D^\mu\phi+b_2\Big[(D_\mu\phi_\nu)^*D^\mu\phi^\nu+\frac{1}{2}(D_\mu\tensor{\phi}{_\nu}{^\nu})^* D^\mu\phi+
\frac{1}{2}(D_\mu\phi)^* D^\mu\tensor{\phi}{_\nu}{^\nu}-\nonumber\\
&&-ig(D_\mu\phi)^*\tensor{A}{^\mu}{_\nu}\phi^\nu+
ig\phi_\nu^*\tensor{A}{_\mu}{^\nu}D^\mu\phi-ig(D_\mu\phi_\nu)^*A^{\mu\nu}\phi+ig\phi^*A^{\mu\nu}D_\mu\phi_\nu+\nonumber\\
&&+g^2\phi^*A_{\mu\nu}A^{\mu\nu}\phi-
\frac{1}{2}ig(D_\mu\phi)^*\tensor{A}{^{\mu\nu}}{_\nu}\phi+
\frac{1}{2}ig\phi^*\tensor{A}{^{\mu\nu}}{_\nu}D_\mu\phi
\Big]-U(\Phi)\nonumber\\
\end{eqnarray}
where, 
\begin{equation}
U(\Phi)=\frac{1}{2}\lambda^2\left[
\phi^*\phi+\lambda_2\left(\phi_\mu^*\phi^\mu+\frac{1}{2}\phi^*\tensor{\phi}{_\mu^\mu}+\frac{1}{2}\phi\phi^{*\mu}_{\mu}\right)-\frac{1}{2}\eta^2\right]^2
\end{equation}
and a another two parameter $b_2$, $\lambda_2$ has been introduced, the former being of equal phenomenological importance as $g_2$.

Taking into account \eqref{gauge transformation scalar covariant}-\eqref{gauge transformation tensor covariant} and noting that,
\begin{eqnarray}
\delta(\phi_\mu^*\phi^\mu)&=&ig\xi^\mu\left(\phi_\mu^*\phi-\phi_\mu\phi^*\right)\\
\delta(\phi\phi^{*\mu}_{\mu})&=&-ig\left(2\xi_\mu\phi^{*\mu}\phi+\xi_\mu^{~\mu}\phi^*\phi\right)
\end{eqnarray} 
gauge invariance of $\mathcal{L}_\Phi$ can be verified.

The total Lagrangian under study will, therefore, be
\begin{equation}\label{Total Lagrangian}
\mathcal{L}=\mathcal{L}_B+\mathcal{L}_\Phi
\end{equation}
which can be shown to be invariant under the simultaneous gauge transformations
\eqref{gauge transformations_gauge boson_1}-\eqref{gauge transformations_gauge boson_3} and
\eqref{scalar phi gauge transformation}-\eqref{tensor phi gauge transformation}.

We can get the Noether currents from the total Lagrangian \eqref{Total Lagrangian}, following the standard variational method for the $\Phi$ fields and taking into account their corresponding equations of motion.

\begin{small}
\begin{equation}
J^\mu=\frac{\partial\mathcal{L}}{\partial(\partial_\mu\phi)}\delta\phi+
\frac{\partial\mathcal{L}}{\partial(\partial_\mu\phi^*)}\delta\phi^*+
\frac{\partial\mathcal{L}}{\partial(\partial_\mu\phi_\nu)}\delta\phi_\nu+
\frac{\partial\mathcal{L}}{\partial(\partial_\mu\phi_\nu^*)}\delta\phi_\nu^*+
\frac{\partial\mathcal{L}}{\partial(\partial_\mu\tensor{\phi}{_\nu^\nu})}\delta\tensor{\phi}{_\lambda^\lambda}+
\frac{\partial\mathcal{L}}{\partial(\partial_\mu\phi^{*\nu}_{\nu})}\delta\phi^{*\lambda}_{\lambda}
\end{equation}
\end{small}
We get three independently conserved currents, one for each of the three independent gauge functions $\xi_{\mu}^{~\mu}, \xi_\mu, \xi$.
\begin{eqnarray}
j^\mu_0&=&i\left[\phi (D^\mu\phi)^*-\phi^*D^\mu\phi\right]\label{current_0}\\
j_1^\mu&=&i\Big\{
\phi_\nu(D^\mu\phi^\nu)^*-\phi_\nu^* D^\mu\phi^\nu+
\frac{1}{2}\left[\phi(D_\mu\tensor{\phi}{_\nu}{^\nu})^*-\phi^*D^\mu\tensor{\phi}{_\nu}{^\nu}+\phi_\nu^{~\nu}(D^\mu\phi)^*-\phi_\nu^{*\nu}D^\mu\phi\right]
\nonumber\\
&&\qquad+ig\left[2\left(\phi \phi_\nu^*+\phi^*\phi_\nu\right)A^{\mu\nu}+|\phi|^2\tensor{A}{^{\mu\nu}}{_\nu}\right]\Big\}\\\label{current_1}
j^{\mu\nu}&=&i\left\{\phi(D^\mu\phi^\nu)^*-\phi^*D^\mu\phi^\nu+\phi^\nu(D^\mu\phi)^*-\phi^{*\nu}D^\mu\phi+2ig|\phi|^2A^{\mu\nu}\right\}\label{current_2}
\end{eqnarray}
The explicit form of the Noether currents facilitates the Euler-Lagrange equations. 
In particular, for the vector gauge field $A_\mu$ we get,
\begin{equation}
\frac{\partial\mathcal{L}}{\partial A_\mu}=-g(j^\mu_0+b_2j_1^\mu)
\end{equation}
and the E-L equations are,
\begin{equation}\label{EL_1}
\partial_\mu\left[F^{\mu\nu}+\frac{1}{2}g_2\left(
\tensor{F}{^{\mu\nu,\lambda}_\lambda}+
\tensor{F}{^{\nu\lambda,\mu}_\lambda}+
\tensor{F}{^{\lambda\mu,\nu}_\lambda}
\right)\right]=g(j_0^\nu+b_2j_1^\nu)
\end{equation}
From the above equation it is clear that the vector gauge field couples to the third rank, auxiliary gauge boson. Variation with respect to the 2nd rank tensor gauge field leads to,
\begin{equation}
\frac{\partial\mathcal{L}}{\partial A_{\mu\nu}}=-b_2gj^{\mu\nu}
\end{equation}
and the corresponding E-L equations are,
\begin{equation}\label{EL_2}
g_2\left[\partial_\mu\tensor{F}{^{\mu\nu,\lambda}}
-\frac{1}{2}\left(
\partial_\mu\tensor{F}{^{\mu\lambda,\nu}}+
\partial_\mu\tensor{F}{^{\lambda\nu,\mu}}+
\partial^\lambda\tensor{F}{^{\mu\nu,}_\mu}+
\eta^{\nu\lambda}\partial_\mu\tensor{F}{^{\mu\rho,}_\rho}
\right)\right]
=gb_2j^{\nu\lambda}
\end{equation}
Contrary to the previous case, the equations of motion for the tensor gauge boson are completely decoupled. With the aid of the generalized Bianchi identities, one can also check that the 2nd rank Noether current is conserved with respect to both of its indices \cite{Konitopoulos:2007},
\begin{equation}
\partial_\mu j^{\mu\nu}=\partial_\nu j^{\mu\nu}=0
\end{equation}
Finally, for the third rank auxiliary tensor field\footnote{Since only the trace over its last two indices appears in both the equations of motion and the Lagrangian, the auxiliary field is actually a vector.} we get,
\begin{equation}
\frac{\partial\mathcal{L}}{\partial \tensor{A}{_{\mu\nu}^\nu}}=-\frac{1}{2}b_2gj^{\mu}_0
\end{equation}
and the corresponding E-L equations read,
\begin{equation}\label{EL_3}
g_2\partial_\mu F^{\mu\nu}=2b_2gj_0^\nu
\end{equation}
At this point it is worth to investigate the parameter space of $(g_2, b_2)$ by the way they appear in the equations \eqref{EL_1}, \eqref{EL_2}, \eqref{EL_3}. First of all, if both the parameters are set to zero, \eqref{EL_1} reduces to the standard Maxwell's equation while the \eqref{EL_2} and \eqref{EL_3} vanish identically. This is the standard Abelian-Higgs model limit. 

Allowing the parameters to move along the rest parametric surface,
we see that the auxiliary field act as a Lagrange multiplier, the end result of which is to modify the coupling constant that enters Maxwell's equations. As we shall see in the following lines, this modification is encapsulated in the phenomenological quantities that govern the most striking effects of superconductivity.

To apply the Higgs mechanism in the above Lagrangian, we need to look for static and constant $\Phi$ fields that minimize the potential functional $U(\Phi)$. The vacuum expectation values for the fields $\Phi$ that are compatible with Poincare invariance are accordingly \cite{Savvidy2006},
\begin{equation}
\phi^{(0)}=\frac{\eta}{\sqrt{2}},~\phi^{(0)}_\mu=0,~\phi^{(0)}_{\mu\nu}=0
\end{equation}
Expanding as a fluctuation from the vacuum, 
\begin{equation}
    \phi(x)=\frac{1}{2}\left(\eta+\sigma(x)\right)e^{i\theta(x)}
\end{equation}
the potential functional becomes, 
\begin{equation}
    U(\Phi)=\frac{1}{2}(\lambda\eta)^2\sigma^2+\frac{1}{2}\lambda_2(\lambda\eta)^2\phi_\mu^*\phi^\mu+(\text{interaction terms})
\end{equation}
The masses of the fields $\sigma$ and $\phi_\mu$ can be straightforwardly identified from the above expression as
\begin{eqnarray}
m_\sigma&=&\lambda\eta\\
m_{\phi_\mu}&=&\sqrt{\frac{\lambda_2}{2b_2}}m_\sigma
\end{eqnarray}
The above relations show that the correlation length concerning the $\sigma$ field \eqref{scalar correlation length} does not undergo any modification, but a new correlation length has to be added, for the non-gauge vector field $\phi_\mu$,
\begin{equation}
\xi_c^\prime=\sqrt{\frac{2b_2}{\lambda_2}}\xi_c\
\end{equation}
This measures the distance over which the spin triplet Cooper-pair wave function can vary significantly. 

\section{Higher Spin Extension of London Equations}\label{chapter 4}
In the ground state where all fluctuations are taken to zero, and fixing in the unitary gauge where all would be Goldstone bosons disappear, the covariant derivative of $\phi$ becomes,
\begin{equation}\label{supercovariant}
D_\mu\phi=-\frac{ig}{\sqrt{2}}\eta A_\mu
\end{equation}
Now, the currents \eqref{current_0}-\eqref{current_2} become a set of generalized London equations,
\begin{eqnarray}
j_0^\mu&=&-g\eta^2 A^\mu\\
j_1^\mu&=&-\frac{1}{2}g\eta^2\tensor{A}{^{\mu\nu}}{_\nu}\\
j^{\mu\nu}&=&-g\eta^2A^{\mu\nu}\label{London current_2}
\end{eqnarray}
Plugging the above expressions into \eqref{EL_1}, \eqref{EL_2}, \eqref{EL_3} we get,
\begin{eqnarray}
&&\partial_\mu\left[F^{\mu\nu}+\frac{1}{2}g_2\left(
\tensor{F}{^{\mu\nu,\lambda}_\lambda}+
\tensor{F}{^{\nu\lambda,\mu}_\lambda}+
\tensor{F}{^{\lambda\mu,\nu}_\lambda}
\right)\right]=-g^2\eta^2\left(A^\nu+\frac{1}{2}b_2\tensor{A}{^{\nu\mu}_\mu}\right)\label{London_1}\\
&&g_2\left[\partial_\mu\tensor{F}{^{\mu\nu,\lambda}}
-\frac{1}{2}\left(
\partial_\mu\tensor{F}{^{\mu\lambda,\nu}}+
\partial_\mu\tensor{F}{^{\lambda\nu,\mu}}+
\partial^\lambda\tensor{F}{^{\mu\nu,}_\mu}+
\eta^{\nu\lambda}\partial_\mu\tensor{F}{^{\mu\rho,}_\rho}
\right)\right]
=-b_2\eta^2g^2A^{\nu\lambda}\label{London_2}\\
&&g_2\partial_\mu F^{\mu\nu}=-2b_2g^2\eta^2A^\nu
\label{London_3}
\end{eqnarray}
Examining the parameter space of $(g_2,b_2)$, once again, first of all we see that at the point $g_2=b_2=0$ we get the standard London's equation.
On the line $b_2=\frac{1}{2}g_2$ it is \eqref{London_3} that gives the standard superconductor's equation which means that the auxiliary field $\tensor{A}{_{\mu\nu}^\nu}$ acts as a Lagrange multiplier. Further, plugging \eqref{London_3} in \eqref{London_1}, decouples completely the vector field, thus ending up with three completely decoupled differential equations, one for each of the fields $A_\mu$, $A_{\mu\nu}$, $\tensor{A}{_{\mu\nu}^\nu}$.

In the general case and taking into consideration that the vector gauge boson is proportional to the conserved current and thus, $\partial_\mu A^\mu=0$, equation \eqref{London_3} reduces to a typical Proca equation,
\begin{equation}\label{Proca equation}
\partial_\mu\partial^\mu A^\nu+\tilde{m}_\gamma^2 A^\nu=0
\end{equation}
The mass of the photon has been modified from the standard Ginzburg-Landau scenario to the value,
\begin{equation}
\tilde{m}_\gamma=\sqrt{\frac{2b_2}{g_2}}m_\gamma
\end{equation}
Equation \eqref{Proca equation} can be solved with standard methods, giving an exponential decay for the magnetic field entering the superconductor's area, $B\sim e^{-x/\tilde{\lambda}}$ with a modified London's penetration depth, 
\begin{equation}
    \tilde\lambda_L=\frac{1}{\tilde{m}_\gamma}=\sqrt{\frac{g_2}{2b_2}}\lambda_L
\end{equation}
It is worth pointing out that the tuning value of the parameters on the line $b_2=\frac{1}{2}g_2$ that decouples the three London's equations, sets also the mass of the vector gauge field and therefore the London's penetration depth, to their standard values. 
On the grounds of \eqref{London_3}, \eqref{London_1} becomes
\begin{equation*}
\partial_\mu\left(\tensor{F}{^{\mu\nu,\lambda}_\lambda}+
\tensor{F}{^{\nu\lambda,\mu}_\lambda}+
\tensor{F}{^{\lambda\mu,\nu}_\lambda}\right)=-\frac{2g^2\eta^2}{g_2^2}\left[(g_2-2b_2)A^\nu+\frac{1}{2}g_2b_2\tensor{A}{^{\nu\mu}_\mu}
\right]
\end{equation*}
However, this equation is irrelevant to the current discussion since it merely determines the relation between the vector and the auxiliary, non-dynamic, gauge fields. Lastly, taking into consideration that the 2nd rank tensor gauge boson is proportional to the 2nd rank conserved current \eqref{London current_2} and the fact that this conservation holds for both the indices we get,
\begin{equation}
\partial_\mu A^{\mu\nu}=\partial_\nu A^{\mu\nu}=0
\end{equation}
Then, equation \eqref{London_2} becomes,
\begin{equation}
\partial_\mu\partial^\mu A^{\nu\lambda}-\frac{1}{2}\left[
\partial_\mu\partial^\mu A^{\lambda\nu}-\partial^\lambda\partial^\nu \tensor{A}{^\mu_\mu}
+\eta^{\nu\lambda}\partial_\mu\partial^\mu\tensor{A}{^\rho_\rho}\right]=-m_T^2 A^{\nu\lambda}
\end{equation}
where $m_T=\sqrt{\frac{b_2}{g_2}}m_\gamma$.
If we split $A^{\mu\nu}$ into a traceless symmetric and antisymmetric components\footnote{The trace, irreducible component of the tensor has been gauged away. However, the gauge invariant equation for the trace component is 
\begin{equation*}
    \partial_\mu\partial^\mu A-m_T^2A=0
\end{equation*}
which is obviously tachyonic and a sign that the vacuum chosen is probably metastable. A thorough study of stable vacua at the expense of Lorentz invariance violation is left for the future. 
},
\begin{equation}
A^{\mu\nu}=A_S^{\mu\nu}+A_A^{\mu\nu}
\end{equation}
we get,
\begin{eqnarray}
\label{KG-symmetric}
\partial_\mu\partial^\mu A_S^{\nu\lambda}+2m_T^2A_S^{\nu\lambda}&=&0\\
\label{KG-antisymmetric}
\partial_\mu\partial^\mu A_A^{\nu\lambda}+\frac{2}{3}m_T^2A_A^{\nu\lambda}&=&0
\end{eqnarray}

Equations \eqref{KG-symmetric} and \eqref{KG-antisymmetric} are typical Proca equations that describe the propagation of a massive graviton-like tensor gauge field with 5 propagating modes\footnote{Gravitons are described by traceless, symmetric, 2nd rank tensor gauge fields.} and a massive Kalb-Ramond field with 3 propagating modes \cite{Kalb1974}, respectively. 
Analogously to the case of the vector gauge field, we can define a 2nd rank magnetic tensor by isolating the space components of the 3rd rank field strength and 
dualizing with respect to the first two indices. 
\begin{equation}
B_{ij}=\frac{1}{2}\varepsilon_{ilm}F_{lm,j}=\varepsilon_{ilm}\partial_{l}A_{mj}
\end{equation}
Contracting \eqref{KG-symmetric} and \eqref{KG-antisymmetric} with the Levi-Civita tensor and for static fields we get,
\begin{eqnarray}
&&\nabla^2B_{ij}^S=2m_T^2B_{ij}^S\\
&&\nabla^2B_{ij}^A=\frac{2}{3}m_T^2B_{ij}^A
\end{eqnarray}

Both the above equations admit solutions of the form $B_{ij}\sim e^{-x/\lambda}$, for a superconductor occupying the half space $x>0$. The emergence of the 2nd rank tensor magnetic fields is of purely diamagnetic origin. The fields are created as a collective response to the incoming magnetic field. Higher rank magnetic fields are not penetrating from the outside to the inside of the superconductor, rather they are being created inside and propagate outwards ($x<0$) within the corresponding ranges,
\begin{equation}
\lambda_{S}=\sqrt{\frac{g_2}{2b_2}}\lambda_L~~,~~~
\lambda_{A}=\sqrt{\frac{3g_2}{2b_2}}\lambda_L
\end{equation}

\section{Conclusions}
We extended the relativistic version of Ginzburg–Landau theory, the so-called Abelian Higgs model, to include non-gauge vector fields and second-rank tensor gauge fields. Guided by the gauge principle, the inclusion of auxiliary second-rank non-gauge tensor fields and third-rank tensor gauge fields—both of which act as Lagrange multipliers—was unavoidable. This extension allows for the possibility of forming superconducting condensates other than Cooper pairs in spin-singlet states. The diamagnetic emergence of a massive second-rank tensor gauge field inside the superconductor, together with the auxiliary third-rank tensor gauge field, both modifies the penetration depth of the external magnetic field and determines the ranges of the outgoing tensor magnetic fields. These predictions appear to be experimentally testable and are expected to be fruitful both for understanding extended tensor gauge field theory (including parameter determination) and for improving the accuracy of our description of superconducting behavior.

\nocite{*}
\printbibliography
\end{document}